\documentclass[conference, letter]{IEEEtran}
\IEEEoverridecommandlockouts
\usepackage{cite}
\usepackage{amsmath,amssymb,amsfonts}
\usepackage{epsfig}
\usepackage{theorem}
\usepackage{algorithm}
\usepackage{algorithmic}
\usepackage{graphicx}
\usepackage{textcomp}
\usepackage{xcolor}
\usepackage{mathtools}

\def\BibTeX{{\rm B\kern-.05em{\sc i\kern-.025em b}\kern-.08em
    T\kern-.1667em\lower.7ex\hbox{E}\kern-.125emX}}

\def \C {\mathbb C}

\def \R {\mathbb R}

\newcommand{\wML}{\mathbf{w}_{\text{ML}}}
\newcommand{\wSL}{\mathbf{w}_{\text{SL}}}
\newcommand{\herm}{^{\text{H}}}
\newcommand{\trans}{^{\text{T}}}

\newcommand{\lb}{{\ell}}

\newcommand{\num}{\wSL\trans\vert \mathbf{r} \vert^p}
\newcommand{\den}{\wML\trans \vert \mathbf{r} \vert^p}
\newcommand{\grad}[2]{\nabla_{#1}{#2}}
\begin{document}

\title{Gradient-Descent Based Optimization of Multi-Tone Sinusoidal Frequency Modulated Waveforms
\thanks{David G. Felton's efforts were supported by the Naval Research Enterprise Internship Program (NREIP), and David A. Hague's efforts were supported by the Naval Undersea Warfare Center's In-House Laboratory Independent Research (ILIR) program.}
}

\author{\IEEEauthorblockN{David G. Felton$^1$ David A. Hague$^2$}
\IEEEauthorblockA{\textit{$^1$Radar Systems Lab (RSL), University of Kansas, Lawrence, KS}\\
\textit{$^2$Sensors and Sonar Systems Department, Naval Undersea Warfare Center, Newport, RI}}
}
\maketitle

\begin{abstract}
This paper describes a gradient-descent based optimization algorithm for synthesizing Multi-Tone Sinusoidal Frequency Modulated (MTSFM) waveforms with low Auto-Correlation Function (ACF) sidelobes in a specified region of time delays while preserving the ACF mainlobe width.    The algorithm optimizes the Generalized Integrated Sidelobe Level (GISL) which controls the mainlobe and sidelobe structure of the waveform's ACF.  This optimization is performed subject to nonlinear constraints on the waveform's RMS bandwidth which directly controls the ACF mainlobe width.  Since almost all of the operations of the algorithm utilize the  Fast Fourier Transform (FFT), it is substantially more computationally efficient than previous methods that synthesized MTSFM waveforms with low ACF sidelobes.  The computational efficiency of this new algorithm facilitates the design of larger dimensional and correspondingly larger time-bandwidth product MTSFM waveform designs.  The algorithm is demonstrated through several illustrative MTSFM design examples. 
\end{abstract}


\section{Introduction}
\label{sec:Intro}
The ability to optimize transmit waveforms, known as waveform diversity, has been an active topic of research in the radar community for over two decades \cite{Gini_2012_Waveform, Blunt_Waveform_Diversity}.  This area of research has been enabled by the development of several parameterized modulation techniques such as Phase-Coding (PC) \cite{Levanon} and Frequency Shift-Keying (FSK) \cite{Costas} which facilitate the design of novel waveforms with unique characteristics.  PC waveforms are of particular interest to the radar community, and there exists an extensive collection of computationally efficient algorithms that synthesize PC radar waveforms with desirable correlation properties \cite{Levanon, SoltanalianI, prabhuBabuIII, PalomarII, PrabhuBabuI, prabhuBabuLp2, prabhuBabuLp3, Blunt_PCFM_Grad, JianLiBookII}.  Recently, waveform diversity has become a topic of increasing interest to the active sonar community \cite{Beerens, Costas_Sonar} with diverse sets of waveforms being employed for Multi-Beam Echo Sounding (MBES) \cite{MBES_Orthogonal_Waveforms} and a variety of Multiple-Input Multiple Output (MIMO) sonar applications \cite{mimoWaveformII, MIMO_SAS, Hansen_MIMO_SAS, mimoWaveformIII}.  These efforts highlight the need for continued development of parameterized waveform models for active sonar waveform diversity.  

Recently, the Multi-Tone Sinusoidal Frequency Modulated (MTSFM) waveform was introduced as a novel FM-based parameterized waveform model.  The MTSFM waveform's phase/frequency modulation functions are composed of a finite Fourier series.  The Fourier coefficients representing the waveform's instantaneous phase are utilized as a discrete set of adjustable parameters \cite{Hague_ASA_2022, Hague_IEEE_AES}.  Previous efforts have demonstrated that the MTSFM coefficients can be modified to shape the mainlobe and sidelobe structure of the waveform's Ambiguity Function (AF) and Auto Correlation Function (ACF) \cite{Hague_IEEE_AES, Hague_OCEANS_2022, Hague_SSP_2018}.  One uniquely advantageous characteristic of the MTSFM waveform is that its spectrum is much more highly concentrated in its swept band of frequencies than PC and FSK waveforms \cite{Hague_IEEE_AES}.  As such, the MTSFM waveform is ideally suited for efficient transmission on practical piezoelectric sonar projectors \cite{Hague_JASA_2019}.  The adaptability of the MTSFM coupled with its natural constant envelope and spectral compactness properties makes it an excellent waveform for practical use in a variety of active sonar systems.  

Currently, the primary design challenge for the MTSFM waveform model is the development of computationally efficient algorithms that synthesize MTSFM waveforms with desirable characteristics.  The majority of the aforementioned efforts in designing MTSFM waveforms with low AF/ACF sidelobes \cite{Hague_IEEE_AES, Hague_OCEANS_2022} utilized optimization routines using the MATLAB\textsuperscript{\textregistered}~optimization toolbox, namely the \emph{fmincon} function \cite{Matlab}.  The algorithms optimized the MTSFM's AF/ACF sidelobes via a $\ell_p$-norm metric on the sidelobes over sub-regions in the range-Doppler plane similar to the algorithms developed in \cite{prabhuBabuLp2, prabhuBabuLp3}.  

While the aforementioned MTSFM optimization algorithms were versatile and highly effective in that they can uniquely shape the sidelobe structure of the AF/ACF, they are not streamlined to be extremely computationally efficient.  The most computationally efficient version of these algorithms used an interior-point method.  One of the primary steps in this interior-point method performs a modified Cholesky decomposition on the Hessian of the waveform design objective function at each iteration \cite{fmincon}.  This is the most computationally expensive step of the algorithm \cite{fmincon} and is particularly burdensome for large dimensional problems since the size of the Hessian grows as the square of the dimensionality $L$ of the problem.  Since this dimensionality $L$ is a proper fraction of the waveform's Time-Bandwidth Product (TBP), the computational bottleneck of this method has correspondingly limited its application to small TBP waveform designs \cite{Hague_IEEE_AES}.  A recent result in \cite{Hague_Asilomar_2022} developed a structured phase retrieval algorithm loosely based off of efforts in \cite{Jian_Li_CAN} that synthesizes MTSFM waveforms with low ACF sidelobes in a substantially more computationally efficient manner than the aforementioned algorithms in \cite{Hague_IEEE_AES}.  This cyclic algorithm specifically optimizes the MTSFM's ACF via a $\ell_2$-norm metric on the ACF sidelobes over all time lags.  It is not, however, capable of reducing sidelobes over sub-regions in time-delay as previous MTSFM optimization techniques did in \cite{Hague_IEEE_AES, Hague_OCEANS_2022}, nor can it optimize more general $\ell_p$-norm metrics like those in \cite{Hague_IEEE_AES, prabhuBabuLp2, prabhuBabuLp3}.  

This paper introduces a gradient-descent based algorithm that synthesizes MTSFM waveforms with low ACF sidelobes in a specified sub-region of time delays via minimization of a more general $\ell_p$-norm metric known as the Generalized Integrated Sidelobe Level (GISL).  The algorithm leverages methods developed in \cite{Blunt_PCFM_Grad} that were used to optimize Polyphase-Coded FM (PCFM) waveforms and, more recently, Constant-Envelope Orthogonal Frequency Division Mutliplexing waveforms \cite{Hague_Felton_CE_OFDM_2} utilized in Dual Function Radar-Communications (DFRC) applications.   Since this Gradient-Descent GISL (GD-GISL) algorithm's operations are largely composed of FFTs, it is computationally efficient, which facilitates synthesizing large-dimensional waveform design problems.  In addition to its computational efficiency, it is more versatile in that it optimizes over the more general $\ell_p$-norm metric for sub-regions of time delays like the algorithms in \cite{Hague_IEEE_AES, prabhuBabuLp2, prabhuBabuLp3}.  Several illustrative design examples demonstrate GD-GISL algorithm's ability to finely tune the sidelobe structure of the MTSFM waveform's ACF and are readily scalable to much larger dimensional problems and therefore larger TBP MTSFM waveform designs than the previous efforts in \cite{Hague_IEEE_AES}.  The rest of this paper is organized as follows: Section II descibes the MTSFM waveform model and the design metrics used to optimize its ACF characteristics; Section III describes the GD-GISL algorithm; Section IV evaluates the performance of this algorithm via several illustrative design examples; finally, Section V concludes the paper. 

\section{MTSFM Waveform Design}
\label{sec:sigModel}
The general FM waveform model is expressed in the time domain as
\begin{equation} \label{eq:sig_mod}
s\left(t\right)=a\left(t\right) e^{j\varphi(t)}e^{j2\pi f_c t},~-\frac{T}{2}\leq t \leq \frac{T}{2}
\end{equation}
where $a\left(t\right)$ is a real and positive amplitude tapering function, $\varphi\left(t\right)$ is the waveform's phase modulation function, $T$ is the waveform's duration, and $f_c$ its carrier frequency.  This paper assumes that the waveform is normalized to unit-energy and basebanded to DC (i.e., $f_c = 0$).  The MTSFM waveform's instantaneous phase is expressed as a finite Fourier series \cite{Hague_IEEE_AES}
\begin{equation}
\varphi(t)= \dfrac{\alpha_0}{2} + \sum_{\ell=1}^L \alpha_{\ell}\cos\left(\dfrac{2\pi \ell t}{T}\right) + \beta_{\ell}\sin\left(\dfrac{2\pi \ell t}{T}\right)
\label{eq:mtsfmPhase}
\end{equation}
where $L$ is the number of Fourier series harmonics in the waveform's instantaneous phase, $\alpha_0$ is a constant phase term, and $\alpha_{\ell}$ and $\beta_{\ell}$ are the waveform's modulation indices.  The modulation indices form a discrete set of $2L$ parameters that are modified to synthesize MTSFM waveforms with desirable ACF properties.  The waveform's corresponding frequency modulation function $m\left(t\right)$ is expressed as
\begin{align}
m\left(t\right) &= \frac{1}{2\pi}\frac{\partial\varphi\left(t\right)}{\partial t} \IEEEnonumber\\ &= \sum_{\ell=1}^L \left(\dfrac{-\alpha_{\ell}\ell}{T}\right)\sin\left(\dfrac{2\pi \ell t}{T}\right) + \left(\dfrac{\beta_{\ell}\ell}{T}\right)\cos\left(\dfrac{2\pi \ell t}{T}\right).
\label{eq:modFunc}
\end{align}
Since the MTSFM waveform's phase modulation function is expressed as a finite Fourier series, it is infinitely differentiable \cite{boyd}.  This property makes these functions smooth and devoid of any transient components.  This results in the vast majority of the MTSFM waveform's spectral content being densely concentrated in a compact band of frequencies.  Coupling this spectral compactness property with its natural constant envelope makes the MTSFM waveform model ideally suited for efficient transmission on piezoelectric sonar transmitters. 

Assuming a narrowband Doppler model, the AF measures the waveform's matched filter response to Doppler shifted versions of the transmit waveform and is expressed as
\begin{equation}
\chi\left(\tau, \nu\right) = \int_{-\infty}^{\infty} s\left(t\right)s^*\left(t+\tau\right)e^{j2\pi\nu t} dt
\label{eq:AF}
\end{equation}
where $\nu$ is the Doppler frequency shift expressed as
\begin{equation}
\nu = \dfrac{2\dot{r}}{c_s}f_c
\end{equation}
where $\dot{r}$ is range rate of the target's echo and $c_s$ is the speed of sound in the underwater acoustic medium.  The zero-Doppler cut of the AF (i.e., when $\nu=0$), the ACF, provides the range response of the waveform's MF output and is expressed as
\begin{equation}
R\left(\tau\right) = \chi\left(\tau, \nu\right)|_{\nu=0} = \int_{-\infty}^{\infty} s\left(t\right)s^*\left(t+\tau\right) dt.
\label{eq:ACF}
\end{equation}

There are several metrics that describe the sidelobe structure of a waveform's ACF.  One metric that has found extensive use in waveform optimization is the GISL \cite{Blunt_Waveform_Diversity}.  The GISL evaluates the ratio of $\ell_p$-norms \cite{PrabhuBabuI, Blunt_PCFM_Grad} of the sidelobe and mainlobe regions of the ACF and is expressed as
\begin{IEEEeqnarray}{rCl}
\text{GISL}~= \left(\dfrac{\int_{\Omega_{\tau}}\left|R\left(\tau\right)\right|^p d\tau}{\int_{0}^{\Delta \tau}\left|R\left(\tau\right)\right|^p d\tau}\right)^{2/p}
\label{eq:GISL}
\end{IEEEeqnarray} 
where $p \geq 2$ is an integer and $\Delta \tau$ is the first null of the ACF which in turn denotes the mainlobe width of the ACF as $2\Delta\tau$.  The $\Omega_{\tau}$ term represents a sub-region of time delays excluding the mainlobe region.  When $p=2$, the GISL becomes the standard ISL metric which is often used in radar waveform design \cite{Blunt_PCFM_Grad}.  As $p\rightarrow\infty$, the integrals in \eqref{eq:GISL} approach the infinity norm $||\cdot||_{\infty}^2$, also known as the maximum norm.  Taking the maximum of the mainlobe and sidelobe region simplifies the GISL metric to the Peak-to-Sidelobe Level Ratio (PSLR) metric \cite{Blunt_PCFM_Grad}.  For waveform optimization applications, the maximum norm tends to produce a discontinuous objective function which prevents the efficient use of gradient-descent based waveform optimization methods.  Making $p$ large but finite \cite{Blunt_PCFM_Grad, prabhuBabuLp2, prabhuBabuLp3} results in a smooth objective function that approximates the PSLR metric and is efficiently traversed using gradient-descent based optimization methods. 

\section{The Gradient-Based GISL Algorithm}
\label{sec:GISL}
The design objective of this paper is to develop an algorithm that reduces the MTSFM waveform's ACF sidelobes via the GISL metric while largely preserving its mainlobe width which determines range resolution.  One effective method of ensuring the mainlobe width stays largely fixed is to place a design constraint on the waveform's RMS bandwidth $\beta_{rms}^2$ expressed as \cite{Ricker, Rihaczek}
\begin{equation}
\beta_{rms}^2 = \int_{\infty}^{\infty}\left(f - f_0\right)^2 |S\left(f\right)|^2 df
\end{equation}
where $f_0$ is the waveform's spectral centroid and $S\left(f\right)$ is the waveform's spectrum.  The inverse of the RMS bandwidth accurately approximates the area under the mainlobe of the ACF (i.e., the denominator of \eqref{eq:GISL}) for the case when $p=2$ \cite{Rihaczek}.  As such, placing a constraint on the RMS bandwidth directly translates to constraining the area under the ACF mainlobe.  This directly corresponds to preserving the ACF mainlobe width and therefore the waveform's range resolution.  Conveniently, the MTSFM waveform's RMS bandwidth is expressed in exact closed form as a function of the modulation indices \cite{Hague_SSP_2018}
\begin{equation}
\beta_{rms}^2 = \left(\frac{2\pi}{T}\right)^2 \sum_{\ell=1}^L \ell^2 \left(\dfrac{\alpha_{\ell}^2 + \beta_{\ell}^2}{2}\right).
\label{eq:rmsBand}
\end{equation}
Formally, the optimization problem for reducing the GISL subject to constraints on the RMS bandwidth $\beta_{rms}^2$ is stated as
\begin{align}\underset{\alpha_{\ell},\beta_{\ell}}{\text{min}}~&\text{GISL}\left(\{\alpha_{\ell}, \beta_{\ell}\}, p\right) \IEEEnonumber  \\ \text{s.t. }  &\beta_{rms}^2\left(\{\alpha_{\ell},\beta_{\ell}\}\right) \leq \left(1+\delta\right)\beta_{rms}^2\left(\{\alpha_{\ell}^{\left(0\right)},\beta_{\ell}^{\left(0\right)}\}\right) \IEEEnonumber \\ &\beta_{rms}^2\left(\{\alpha_{\ell},\beta_{\ell}\}\right) \geq \left(1-\delta\right)\beta_{rms}^2\left(\{\alpha_{\ell}^{\left(0\right)},\beta_{\ell}^{\left(0\right)}\}\right)
\label{eq:Prob1}
\end{align}
where $\beta_{rms}^2\left(\{\alpha_{\ell}^{\left(0\right)},\beta_{\ell}^{\left(0\right)}\}\right)$ denotes the initialized waveform's RMS bandwidth (i.e., at iteration $i=0$) and $\delta$ is a unitless bound parameter.  The rest of this section describes the GD-GISL algorithm that solves \eqref{eq:Prob1}.

We attempt to solve the optimization problem stated in \eqref{eq:Prob1} by restating it as an unconstrained optimization problem where the nonlinear inequality constraints are expressed as quadratic penalty functions \cite{numericalOpt}.  This new objective function is expressed as
\begin{equation}
Q\left(\phi_{\ell}, p, \gamma\right) = \text{GISL}\left(\phi_{\ell}, p\right)+\frac{\gamma}{2}\sum_{k\in\mathcal{K}} \left(\left[c_k\left(\phi_{\ell}\right)\right]^- \right)^2
\label{eq:Prob1_Pen}
\end{equation} 
where $\phi_{\ell} = \{\alpha_{\ell}, \beta_{\ell}\}$, $\gamma$ is a unitless penalty parameter, the $\left[x\right]^-$ operator denotes $\text{max}\{-x, 0\}$, and $c_k\left(\phi_{\ell}\right)$ represents the $\mathcal{K}=2$ nonlinear constraint functions which are now expressed as 
\begin{align}
c_1\left(\phi_{\ell}\right):\beta_{rms}^2\left(\phi_{\ell}\right)-\left(1+\delta\right)\beta_{rms}^2\left(\phi_{\ell}^{\left(0\right)}\right) &\leq 0 \label{eq:c1} \\  c_2\left(\phi_{\ell}\right):\left(1-\delta\right)\beta_{rms}^2\left(\phi_{\ell}^{\left(0\right)}\right) -  \beta_{rms}^2\left(\phi_{\ell}\right) &\leq 0.
\label{eq:c2}
\end{align}
The purpose of the penalty functions \eqref{eq:c1} and \eqref{eq:c2} are to substantially increase the objective function via $\gamma$ for values of $\alpha_{\ell}$ and $\beta_{\ell}$ outside the feasible region as specified by the nonlinear equality constraints in \eqref{eq:Prob1}.  As a result of this, any set of values for $\alpha_{\ell}$ and $\beta_{\ell}$ outside this feasible region will produce a large objective function and, most likely, a large positive gradient.  A gradient descent algorithm will compute a search direction away from these increasing values thus ensuring the nonlinear inequality constraints are enforced. 

The first step in developing the gradient-based GISL optimization algorithm is to discretize the waveform signal model and its design metrics.  The MTSFM waveform's instantaneous phase \eqref{eq:mtsfmPhase} can be written as a linear sum using discrete variables as
\begin{equation}
	\mathbf{\varphi} = \begin{bmatrix}
		\mathbf{B}_\mathrm{c} & \mathbf{B}_\mathrm{s}
	\end{bmatrix} \begin{bmatrix}
	\boldsymbol\alpha \\ \boldsymbol\beta
\end{bmatrix} 
= \mathbf{B}\boldsymbol\phi
\end{equation} 
where $\boldsymbol\phi =\left[\boldsymbol\alpha,  \boldsymbol\beta\right]^{\text{T}} = \left[\alpha_1, \alpha_2, \dots, \alpha_L, \beta_1, \beta_2, \dots, \beta_L \right]\trans$ is a $2L\times1$ vector containing the MTSFM's modulation indices and $\mathbf{B}$ is a $M \times 2L$ concatenation of the $ M{\times}L $ basis matrices $\mathbf{B}_\mathrm{c}$ and $\mathbf{B}_\mathrm{s}$ which contain cosine and sine harmonics, respectively, such that the $ \ell^{\text{th}} $ columns
\begin{align}
\mathbf{b}_{\mathrm{c}, \lb} &= \cos\left(\dfrac{2\pi \lb t}{T} \right), \\
\mathbf{b}_{\mathrm{s}, \lb} &= \sin\left(\dfrac{2\pi \lb t}{T} \right)
\end{align}
are sampled at a sampling rate $f_s$ that satisfies the Nyquist criterion.  The Fourier basis used to describe the MTSFM's instantaneous phase is one of several bases that have been utilized for gradient-based optimization \cite{Blunt_PCFM_Grad, Blunt_Alternative_Bases, Hague_Felton_CE_OFDM_2}.  The primary difference in the GD-GISL algorithm described here is that it is optimizing the GISL with RMS bandwidth penalty terms as seen in \eqref{eq:Prob1_Pen} for the MTSFM waveform model.  It's also worth noting that the MTSFM can be implemented with an instantaneous phase that uses the full cosine and sine harmonic basis $\mathbf{B}$ or just $\mathbf{B}_\mathrm{c}$ or $\mathbf{B}_\mathrm{s}$ separately.  Using solely even or odd harmonics in the phase/frequency modulation functions influences the shape of the waveform's resulting AF as is described in \cite{Hague_IEEE_AES, Hague_SSP_2018}.  An additional advantage of doing this is that the dimensionality of the optimization problem is reduced from $2L$ to $L$ which allows for faster convergence to a solution of the optimization problem.  

From here, the development of the GD-GISL algorithm largely follows the descriptions given in \cite{Blunt_PCFM_Grad, Hague_Felton_CE_OFDM_2}.  The GISL metric can be expressed in terms of the discretized ACF which is expressed as
\begin{equation} \label{eq:discr}
\mathbf{r} = \mathbf{A}\herm \lvert \mathbf{A\bar{s}} \rvert^2
\end{equation}
where $\mathbf{r} \in \C^{(2M-1)}$  contains discretized samples of the ACF, $\mathbf{\bar{s}} \in \C^{(2M-1)}$ is a discretized and zero-padded version of $\mathbf{s}$, and $\mathbf{A}$ and $\mathbf{A}\herm$ are $2M-1 \times 2M-1$ Discrete Fourier Transform (DFT) and Inverse DFT matrices, respectively. The vectors $\wSL$ and $\wML$ $\in \R^{(2M-1)}$ are zero everywhere except in the extent of the sidelobe and mainlobe regions, respectively.  The GISL metric is then expressed as the cost function
\begin{equation}
\text{GISL}\left(\boldsymbol\phi, p \right) = \dfrac{\|\wSL \odot \mathbf{r} \|_p^2}{\|\wML \odot \mathbf{r} \|_p^2}.
\label{eq:GISLmyNISL}
\end{equation}

The new unconstrained optimization problem can be formally stated as
\begin{equation}
\underset{\boldsymbol\phi}{\text{min}}~Q\left(\boldsymbol\phi, p, \gamma\right).
\label{eq:Prob2}
\end{equation}
The GISL for the MTSFM waveform is a  $2L$-dimensional and highly non-convex objective function across the MTSFM parameter space $\boldsymbol\phi$.  Therefore, convergence to the global minimum is almost certainly not guaranteed.  We traverse this non-convex objective function using gradient descent.  Gradient descent is an iterative approach which takes some step $\mu$ in the direction of steepest descent $\mathbf{q}_i$
\begin{align}
\boldsymbol\phi_{i+1} &= \boldsymbol\phi_i + \mu \mathbf{q}_i  \label{eq:phi_i} \\
\mathbf{q}_i &= -\mathbf{\nabla}_{\boldsymbol\phi}Q\left(\boldsymbol\phi_i, p, \gamma\right) \label{eq:qi} 
\end{align} 
where $\nabla_{\boldsymbol\phi}$ is the gradient operator.  The gradient of \eqref{eq:Prob2} is expressed as
\begin{equation}
\nabla_{\boldsymbol\phi}Q\left(\boldsymbol\phi, p, \gamma\right) = \nabla_{\boldsymbol\phi}\text{GISL}\left(\boldsymbol\phi, p \right) + \gamma\sum_{k\in\mathcal{K}}c_k\left(\boldsymbol\phi\right)\nabla_{\boldsymbol\phi}c_k\left(\boldsymbol\phi\right)
\label{eq:grad1}
\end{equation}
where $\nabla_{\boldsymbol\phi}Q\left(\boldsymbol\phi, p, \gamma\right)$ is expressed in vector form \cite{Hague_Felton_CE_OFDM_2} as
\begin{equation}
	\grad{\boldsymbol\phi}{J_p}=4 Q\left(\boldsymbol\phi, p, \gamma\right) \mathbf{\bar{D}}^T \Im{\Biggl\{\mathbf{\bar{s}}^* \odot \mathbf{A}\herm \left[(\mathbf{A\bar{s}}) \odot \mathbf{P} \right] \Biggr\}} 
\label{eq:gradDevil1}
\end{equation}
and
\begin{equation}
\mathbf{P} = \Re\left\{\mathbf{A} \left( \lvert\mathbf{r}\rvert^{p-2}\odot\mathbf{r}\odot \left[ \frac{\wSL}{\num} - \frac{\wML}{\den} \right]\right)\right\}.
\label{eq:gradDevil2}
\end{equation}
Typically, performing \eqref{eq:phi_i} and \eqref{eq:qi} iteratively until the Euclidean length of $\mathbf{\nabla}_{\boldsymbol\phi}Q\left(\boldsymbol\phi_i, p, \gamma\right)$ is below some threshold $g_{\text{min}}$ ensures that $Q\left(\boldsymbol\phi_i, p, \gamma\right)$ is very near a local minimum.  We observed empirically that a better stopping criteria for this algorithm is when the Euclidean norm between $\nabla_{\boldsymbol\phi}Q\left(\boldsymbol\phi_i,p,\gamma\right)$ and $\nabla_{\boldsymbol\phi}Q\left(\boldsymbol\phi_{i-1},p,\gamma\right)$ is below the threshold $g_{\text{min}}$.  This tends to prevent the algorithm from running additional iterations that do not produce a substantial improvement in the reduction of the objection function in \eqref{eq:Prob2}.  Alternatively, the routine may continue until it reaches a predetermined number of iterations $I_{\text{max}}$.

We employ heavy-ball gradient descent which includes weighted versions of the previous search directions with the current gradient.  This has been shown to converge quickly for these types of problems by dampening rapid transitions of the gradient thereby enforcing a smooth path to the minima. The search direction is altered by inclusion of previous gradients as
\begin{equation}
\mathbf{q}_i = -\nabla_{\boldsymbol\phi}Q\left(\boldsymbol\phi_i, p, \gamma\right) + \beta \mathbf{q}_{i-1}
\end{equation}
where $\beta \in \left[0, 1\right]$.  Since this method does not always ensure a descent, if in fact the current search direction is an ascent (i.e., the projection of the gradient onto the current search direction is positive), the current search direction is reset to the current gradient.
\begin{equation}
\text{if}~\mathbf{q}_i\trans(\nabla_{\boldsymbol\phi}Q\left(\boldsymbol\phi_i, p, \gamma\right) )> 0,~\text{then}~\mathbf{q}_i = -\nabla_{\boldsymbol\phi}Q\left(\boldsymbol\phi_i, p, \gamma\right).
\end{equation}
Once the search direction is established, a simple backtracking method is used to calculate the step size $\mu$ for the line search that satisfies sufficient decrease via the Armijo condition \cite{numericalOpt}.  The steps of the GD-GISL algorithm are listed in Algorithm 1.  Since the algorithm makes extensive use of FFTs in computing the GISL metric \eqref{eq:gradDevil1}, it is likely to be substantially more computationally efficient than the legacy MTSFM optimization algorithms in \cite{Hague_IEEE_AES}.

\begin{algorithm}
 \caption{The Gradient-Based GISL Algorithm}
 \begin{algorithmic}[1]
 \renewcommand{\algorithmicrequire}{\textbf{Input:}}
 \renewcommand{\algorithmicensure}{\textbf{Output:}}
 \REQUIRE Initialize $\mathbf{B}$, $\mathbf{\phi}^{(0)}$, $P$, $L$, $\mathbf{q}_0 = \mathbf{0}_{\text{N}\times1}$, $\beta$, $\mu$, $\rho_{\text{up}}$, $\rho_{\text{down}}$, $\delta$, $\gamma$, $c$, $g_{\text{min}}$, $I_{\text{max}}$, and set $i=1$.
 \ENSURE  Final MTSFM coefficient vector $\boldsymbol\phi$ with refined ACF properties that locally solves the criteria in \eqref{eq:Prob2}
  \STATE Evaluate $Q\left(\boldsymbol\phi_{i-1}, p, \gamma\right)$ and $\nabla_{\boldsymbol\phi}Q\left(\boldsymbol\phi_{i-1}, p, \gamma\right)$ via \eqref{eq:Prob1_Pen} and \eqref{eq:grad1}.
  \STATE $\mathbf{q}_i = -\nabla_{\boldsymbol\phi}Q\left(\boldsymbol\phi_{i}, p, \gamma\right) + \beta \mathbf{q}_{i-1}$
  \STATE \textbf{If} $\left(\nabla_{\boldsymbol\phi}Q\left(\boldsymbol\phi_{i-1}, p, \gamma\right) \right)^\text{T}~\mathbf{q}_i \geq 0$
  \STATE ~~~~$\mathbf{q}_i = -\nabla_{\boldsymbol\phi}Q\left(\boldsymbol\phi_{i-1}, p, \gamma\right)$
  \STATE \textbf{End}(If)
  \STATE \textbf{While} \begin{multline} Q\left(\boldsymbol\phi_i{+}\mu\mathbf{q}_i, p, \gamma\right)  \label{eq:xi}){>}Q(\boldsymbol\phi_{i-1}, p, \gamma) \\ +c\mu\left(\nabla_{\boldsymbol\phi}Q\left(\boldsymbol\phi_{i-1},p,\gamma\right) \right)^\text{T}\mathbf{q}_i,~\mu= \rho_{\text{down}}\mu\end{multline}
  \textbf{End}(While)
  \STATE $\boldsymbol\phi_i = \boldsymbol\phi_{i-1} + \mu\mathbf{q}_i,~~\mu= \rho_{\text{up}}\mu$
  \STATE $i = i+1$
  \STATE Repeat steps 1-8 until $i=I_{\text{max}}$ or \\ $\|\nabla_{\boldsymbol\phi}Q\left(\boldsymbol\phi_i,p,\gamma\right) - \nabla_{\boldsymbol\phi}Q\left(\boldsymbol\phi_{i-1},p,\gamma\right) \|_2 \leq g_{\text{min}}$
 \end{algorithmic} 
 \end{algorithm}

\section{Several Illustrative Design Examples}
\label{sec:Examples}
This section demonstrates the GD-GISL algorithm using several MTSFM waveform optimization design examples.  In each example, the waveform time series is sampled at a rate $f_s = 10\Delta f$ where $\Delta f$ is the waveform's swept bandwidth.  The time series is also tapered with a Tukey window with shape parameter $\alpha_T = 0.05$ \cite{Harris}.  The algorithm parameters used for each example are shown in Table \ref{table:MTSFM_I}.  All examples were run on a HP EliteBook 845 G8 with a 2.3 GHz AMD Ryzen PRO 565OU processor and 16 GB DDR3 RAM running MATLAB\textsuperscript{\textregistered} version R2019a.  Each design example in this paper uses only the sine basis $\mathbf{B}_\mathrm{s}$ to represent the waveform's instantaneous phase (i.e., only $\beta_{\ell}$ are non-zero).  This produces a waveform with a frequency modulation function, as shown in \eqref{eq:modFunc}, that is even symmetric.  This results in a waveform and with a ``Thumbtack-Like'' AF shape \cite{Rihaczek} that possesses a distinct mainlobe at the origin whose widths in range and Doppler are inversely proprotional to the waveform's bandwidth and duration respectively.  Additionally, this AF shape possesses a pedestal of sidelobes whose height is inversely proportional the waveform's TBP.  While other AF shapes are possible with the MTSFM waveform, the ``Thumbtack-Like'' AF shape was chosen due to ease of implementation and for illustrative purposes.  It is much easier to compare and visualize the reduction in sidelobe levels of the optimized waveform when the seed waveform's sidelobes are relatively constant.  Investigating the optimization of MTSFM waveforms with other AF shapes will be the topic of an upcoming paper.

\begin{table}[htb]
\caption[Algorithm parameters used for each MTSFM waveform optimization example.]{Algorithm parameters used for each MTSFM waveform optimization example.}
\label{table:MTSFM_I}
\centering
\begin{tabular}{c c c}\hline
Algorithm Parameter			&		Symbol					&			Value 				\\ \hline
Initial Step Size				&		$\mu$					&			1.0					\\
Sufficient Decrease Parameter	&		$c$						&			0.1 					\\
Forgetting Factor				&		$\beta$					&			0.1					\\
Step Decrease Parameter		&		$\rho_{\text{down}}$	&			0.25				\\
Step Increase Parameter 		&		$\rho_{\text{up}}$		&			1.01				\\
Max Iterations				&		$I_{\text{max}}$			&			500					\\
Gradient Threshold			&		$g_{\text{min}}$			&			$1 \times 10^{-5}$	\\
Penalty Term					&		$\gamma$				&			2					\\
RMS Bandwidth Bound		&		$\delta$					&			0.1					\\ \hline
\end{tabular}
\end{table} 
  
\subsection{Example I : Low TBP with large $p$ over all time-delays}
\label{subsec:exampleI}
The first design example optimizes the MTSFM waveform shown in Figure 1 of \cite{Hague_IEEE_AES}.  This particular MTSFM waveform's instantaneous phase is composed of $L=32$ sine harmonics where the modulation indices $\beta_{\ell}$ take on the values shown in Table 1 of \cite{Hague_IEEE_AES}.  The goal of this optimization problem is to reduce the ACF sidelobes over the region $\Omega_{\tau} \in \Delta\tau \leq |\tau| \leq T$ via the GISL metric where $p = 20$.  Figure 1 shows the ACFs and spectra of the initial seed waveform and the resulting optimized waveform using the GD-GISL algorithm.  The optimized MTSFM waveform's ACF PSLR was reduced from -15.94 dB to -22.62 dB, an overall reduction in PSLR of 6.68 dB.  The RMS bandwidth of the optimized waveform was 1.1021 times larger than the initial seed waveform's RMS bandwidth suggesting that the upper RMS bandwidth nonlinear constraint was active upon completion of the optimization routine.  Of particular importance is the computation time for this example.  The algorithm completed after 113 iterations in only 0.63 seconds.  Running the same optimization routine using the legacy interior-point method used in \cite{Hague_IEEE_AES} completed after 202 iterations with a computation time of 15.48 seconds, roughly 24.5 times longer than the GD-GISL algorithm.  Even for a small dimensional optimization problem, the GD-GISL algorithm is substantially faster than the legacy interior-point algorithm.  

\begin{figure}[!ht]
\centering
\includegraphics[width=0.5\textwidth]{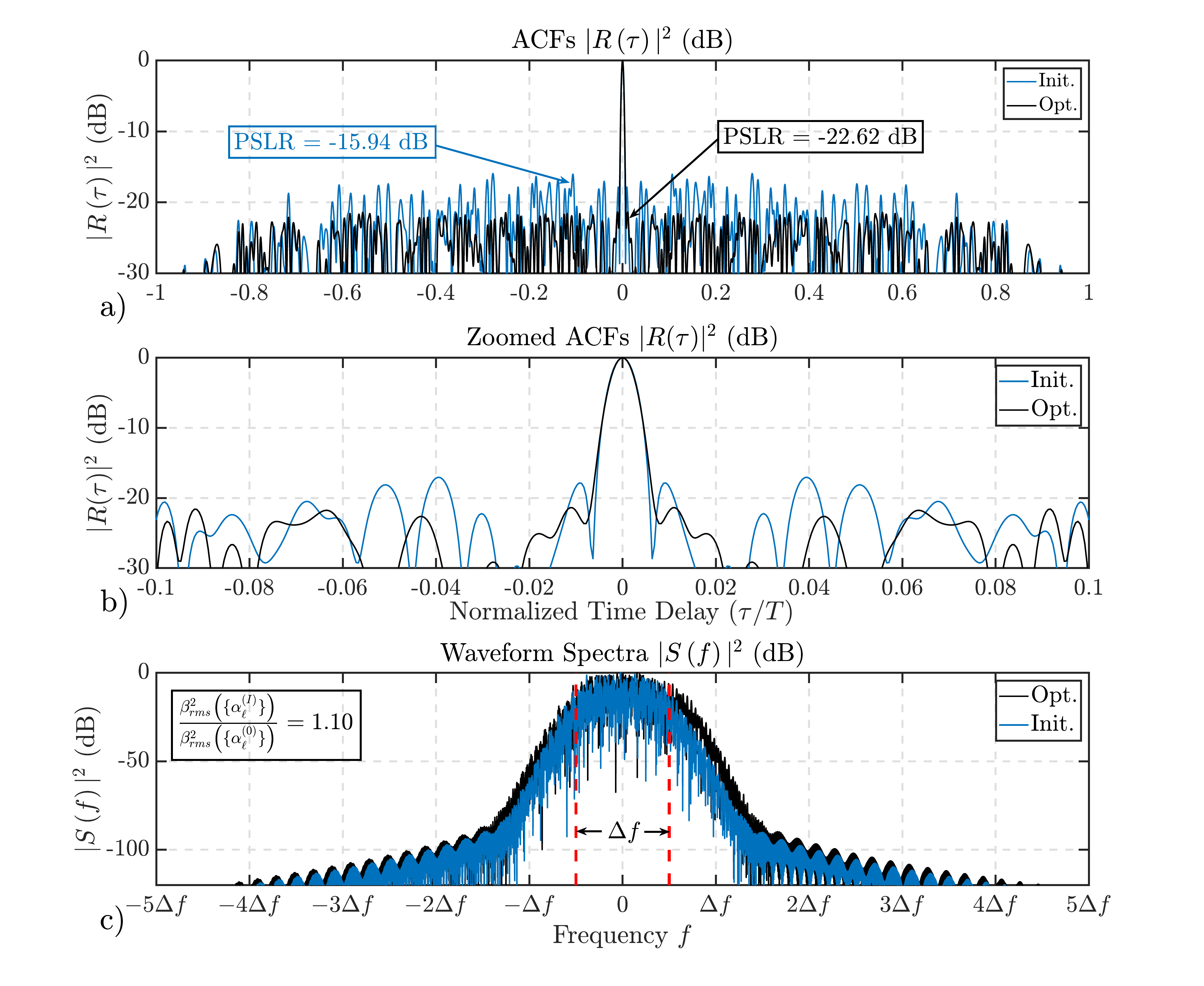}
\caption{ACFs (a), ACFs zoomed at the origin (b), and respective spectra (c) of the initial seed and optimized MTSFM waveforms.  The waveform was optimized over the region $\Omega_{\tau} \in \Delta \tau \leq |\tau| \leq T$ for GISL parameter $p=20$.  The optimized MTSFM waveform possesses clearly lower ACF sidelobes while largely retaining the same mainlobe width.  Correspondingly, the optimized waveform's spectral extent is not substantially different from that of the intial seed waveform.}
\label{fig:MTSFM_1}
\end{figure}

\subsection{Example II : Low TBP with varying $p$ over a sub-region of time-delays}
\label{subsec:exampleII}
\begin{figure}[!ht]
\centering
\includegraphics[width=0.5\textwidth]{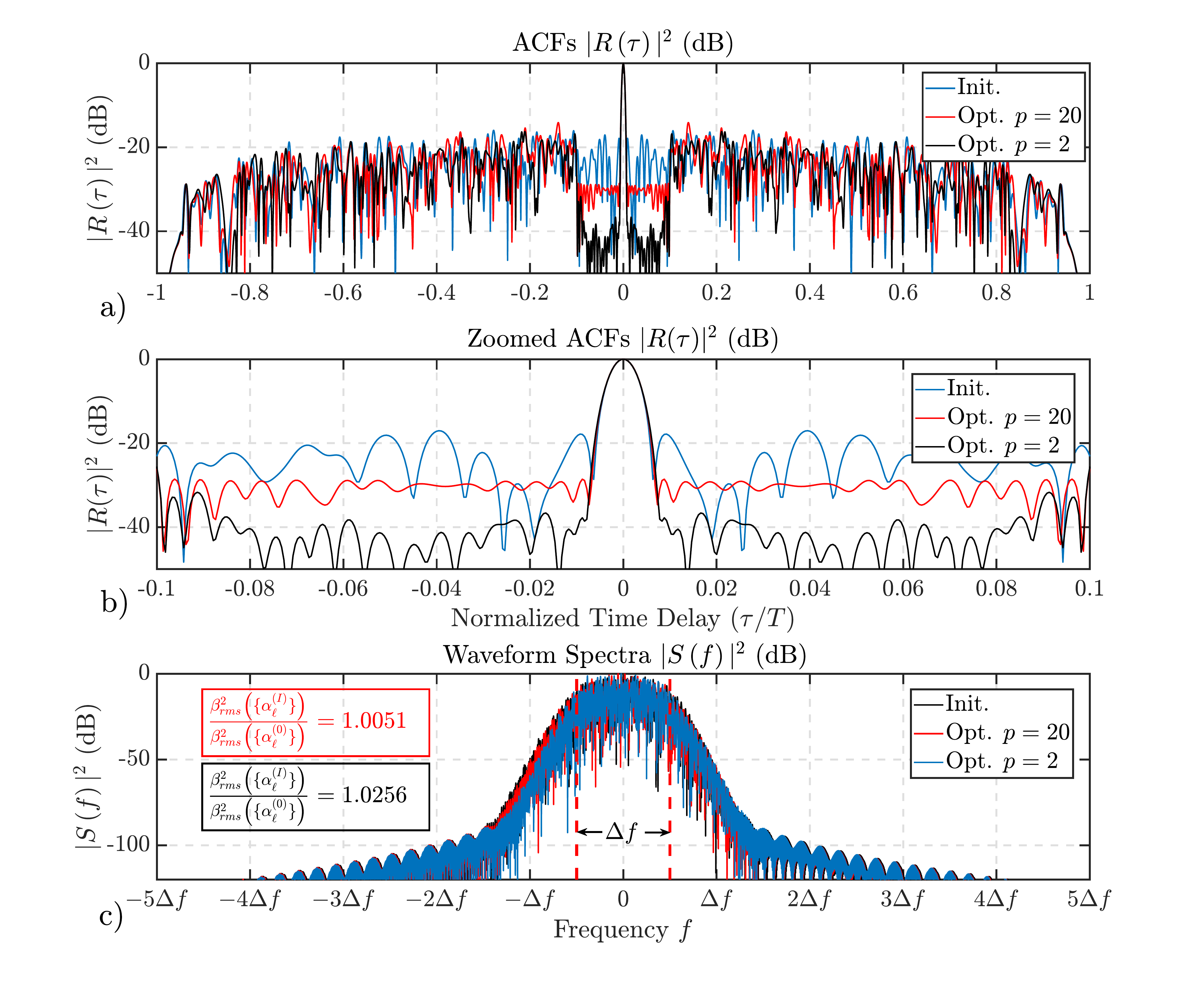}
\caption{ACFs (a), ACFs zoomed at the origin (b), and respective spectra (c) of the initial seed and optimized MTSFM waveforms.  The waveforms were optimized over the region $\Omega_{\tau} \in \Delta \tau \leq |\tau| \leq 0.1T$ for GISL parameters $p=20$ and $p=2$ respectively. The optimized MTSFM waveforms possesses clearly lower ACF sidelobes in $\Omega_{\tau}$ while largely retaining the same mainlobe width.  The sidelobe levels were generally lower for the MTSFM optimized using the GISL metric with $p=2$.  Both optimized waveforms possess essentially the same spectral extent as the intial seed waveform.}
\label{fig:MTSFM_2}
\end{figure}

The second example uses the same initial seed waveform from the previous example but now seeks to optimize the ACF sidelobes over the region $\Omega_{\tau} \in \Delta\tau \leq |\tau| \leq 0.1T$ for the GISL metric using $p = 20$ and then again for $p=2$.  The goal of this example is to demonstrate how the GD-GISL algorithm can finely tune the MTSFM waveform's design coefficients to reduce ACF sidelobes in a very specific region $\Omega_{\tau}$ of time delays.  It also demonstrates the impact that different $p$ values have on the waveform optimization problem.  The ACFs and spectra of the initial seed and optimized waveforms are shown in Figure \ref{fig:MTSFM_2}.  As can be clearly seen from the figure, the ACFs of both optimized waveforms possess substantially lower sidelobes in the region $\Omega_{\tau}$ than the initial seed waveform.  The MTSFM waveform optimized with $p=2$ possesses noticeably lower sidelobes over most of $\Omega_{\tau}$ than the waveform optimized using $p=20$.  Understanding why the GISL metric produces generally lower sidelobes for $p=2$ over $p=20$ likely involves understanding the structure of the GISL objective function with varying $p$.  This will be a topic of future investigation.  As can be seen in panel (c) of Figure \ref{fig:MTSFM_2}, the optimized waveforms' spectra are not substantially altered from that of the initial seed waveform.   Both optimized waveforms' RMS bandwidths were relatively close to that of the seed waveform indicating that the nonlinear RMS bandwidth constraints were not active for either case.  

Both optimization runs completed more quickly than the legacy interior-point algorithm.  For the $p=20$ case, the GD-GISL algorithm completed after 227 iterations in 0.89 seconds.  This was roughly 28.47 times faster than the interior-point method which completed after 244 iterations in 25.34 seconds.  For the $p=2$ case, the GD-GISL algorithm completed after 457 iterations in 1.32 seconds compared to the interior-point method which completed after only 54 iterations in 4.79 seconds.  For this particular case the GD-GISL algorithm, on average, completed an iteration in 0.00289 seconds, and the inter-point method on average completed an iteration every 0.084 seconds, making the GD-GISL algorithm roughly 29 times faster iteration to iteration.  

\subsection{Example III : Large TBP with large $p$ over all time-delays}
\label{subsec:exampleIII}

\begin{figure}[!ht]
\centering
\includegraphics[width=0.5\textwidth]{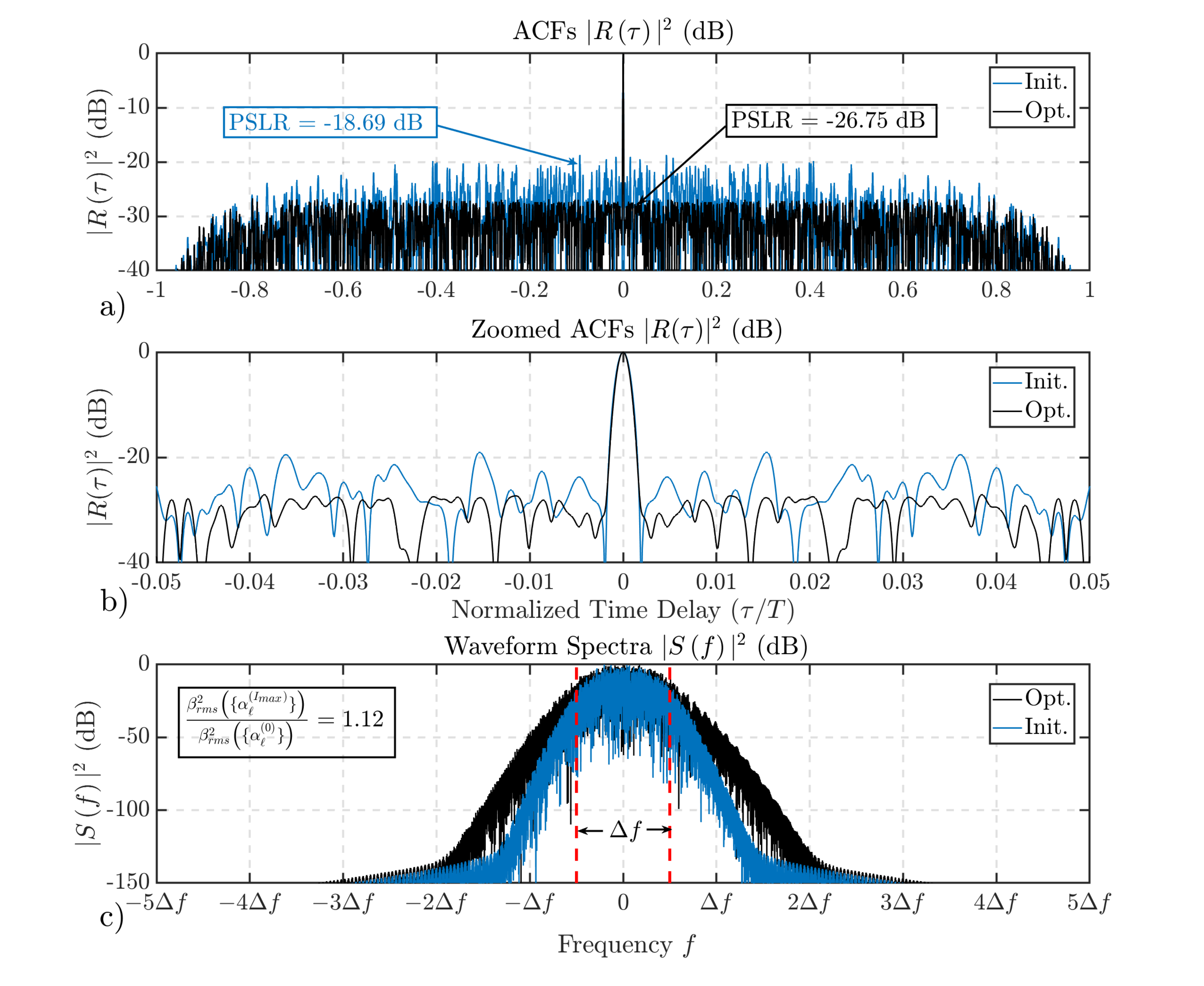}
\caption{ACFs (a), ACFs zoomed at the origin (b), and respective spectra (c) of the initial seed and optimized MTSFM waveforms .  The waveform was optimized over the region $\Omega_{\tau} \in \Delta \tau \leq |\tau| \leq T$ for GISL parameter $p=20$.  The optimized MTSFM waveform possesses clearly lower ACF sidelobe while retaining the same mainlobe width.  Correspondingly, the optimized waveform's spectral extent is not substantially different from the intial seed waveform.}
\label{fig:MTSFM_3}
\end{figure}

The third and final design example demonstrates how the GD-GISL algorithm can handle much larger dimensional problems with relative computational ease compared to the legacy interior-point MTSFM optimization algorithm.  The initial seed MTSFM has a TBP of 1024 and its phase modulation function is composed of $L=256$ sine harmonics.  Based on earlier MTSFM design efforts \cite{Hague_IEEE_AES}, the larger TBP and dimensionality of this design should substantially increase the computation time of the legacy interior-point method.  Figure \ref{fig:MTSFM_3} shows the ACFs and spectra of the initial and GD-GISL ($p=20$) optimized MTSFM waveforms.  The optimized MTSFM waveform's ACF PSLR was reduced to -26.75 dB, a reduction of 8.06 dB from the initial seed waveform.  The RMS bandwidth of the optimized waveform was 1.118 times larger than the initial seed waveform's RMS bandwidth, again suggesting that the upper RMS bandwidth nonlinear constraint was active upon completion of the optimization routine.  Of particular note for this example was the computation time for the GD-GISL algorithm.  It completed after 68 iterations in 10.17 seconds.  A similar optimized MTSFM waveform was achieved using the legacy interior-point algorithm.  It also completed after 68 iterations in 1731.33 seconds, making the GD-GISL algorithm roughly 170 times faster than the legacy interior-point algorithm, a considerable improvement in computational efficiency.    

\section{Conclusion}
\label{sec:Conclusion}
The GD-GISL MTSFM optimization algorithm synthesizes MTSFM waveforms with low ACF sidelobes in a specified sub-region $\Omega_{\tau}$ of time delays via minimization of the GISL metric.  Since most of the algorithm's operations are FFT-based, it is substantially more computationally efficient than the legacy interior-point algorithm used in previous efforts \cite{Hague_IEEE_AES}.  This computational efficiency facilitates synthesizing larger dimensional and consequently larger TBP MTSFM waveform designs in a much shorter amount of time.  The ``Thumbtack-Like'' AF design examples from the last section demonstrated the algorithm's versatility in finely controlling the ACF mainlobe and sidelobe structure as well as its computational efficiency.  Future efforts will focus on extending the versatility and performance of the algorithm in several facets.  The most obvious extension is to design families of MTSFM waveforms with desirably low ACF and Cross-Correlation Function (CCF) sidelobes over user-defined sub-regions of time-delays and varying values for $p$.  Another obvious extension of this algorithm is to modify it to shape the AF sidelobes in a user-defined region of time-delays and Doppler shifts.  From here, marginals of the AF that characterize other waveform design performance characteristics such as the Q-Function \cite{abraham2019} could also be optimized using the same algorithm.   Lastly, the algorithm should readily accommodate addition nonlinear constraints that can finely tune the mainlobe shape of the AF using the model developed in \cite{Hague_SSP_2018}, which will enable the design of Doppler tolerant MTSFM waveforms.


\end{document}